\documentclass[twocolumn,english,prl]{revtex4-1}
\usepackage{color}
\usepackage{graphicx}
\usepackage[T1]{fontenc}
\usepackage{float}
\usepackage{textcomp}
\usepackage{amsmath}
\usepackage{amssymb}
\usepackage{graphicx}
\usepackage{esint}
\usepackage{natbib}
\usepackage{color}

\begin{document}
\title{Nagaoka spin-valley ordering in silicene quantum dots}

\author{Piotr Jurkowski}

\affiliation{AGH University of Science and Technology, Faculty of Physics and
Applied Computer Science,\\
 al. Mickiewicza 30, 30-059 Krak\'ow, Poland}

\author{Bart\l{}omiej Szafran}

\affiliation{AGH University of Science and Technology, Faculty of Physics and
Applied Computer Science,\\
 al. Mickiewicza 30, 30-059 Krak\'ow, Poland}

\begin{abstract}
We study a cluster of quantum dots defined within silicene that hosts confined electron states
with spin and valley degrees of freedom. Atomistic tight-binding and continuum Dirac approximation
are applied for few-electron system 
in quest  for spontaneous valley polarization driven by 
inter-dot tunneling and electron-electron interaction, i.e. a valley counterpart
of itinerary Nagaoka ferromagnetic ordering recently identified in GaAs square cluster of quantum dots
with three excess electrons [P. Dehollain, {\it et al.}, Nature {\bf 579}, 528 (2020)]. 
We find that for Hamiltonian without intrinsic-spin orbit coupling 
the valley polarization in the ground-state can be observed in a range of inter-dot
spacing provided that the spin of the system is frozen by external magnetic field. 
The inter-valley scattering effects are negligible for cluster geometry that supports the valley polarized ground-state. In presence of a strong intrinsic spin-orbit coupling that is characteristic to silicene no external magnetic field is necessary for observation of ground-state that is polarized in both spin and valley. The effective magnetic field
due to the spin-orbit interaction produces a perfect anticorrelation of the spin and valley isospin components in the low-energy spectrum. 
Experimental detection of the spin-valley ground-state polarization by charge response to potential variation is discussed.

\end{abstract}
\maketitle
\section{Introduction}

The Hubbard model for cubic lattice with on-site Coulomb interaction
dominating over the inter-site hopping 
produces spin polarized  ground-state  near half-filling \cite{nagaoka} that is known as
Nagaoka ferromagnetism in the
theory of itinerary ferromagnetism \cite{if}.
Semiconductor quantum dots were pointed out  as a possible two-dimensional realization of the Hubbard
model and artificial molecules or clusters formed by multiple quantum dots
were studied in the context of a spin polarization driven by inter-dot tunneling and electron-electron interaction \cite{square,nielsen}. 
Spin-ordered ground state was recently experimentally identified in electrostatic quantum dots
defined in GaAs \cite{naturexp} in a three-electron system for quantum dots arranged in 
a square cluster, a case previously theoretically studied in Ref. \cite{square}.

In graphene \cite{graphene} and 2D Xenes \cite{xenes} 
the electron states near the charge neutrality 
point are additionally characterized  by the valley isospin due to the presence of two non-equivalent
Dirac points in the Brillouin zone. Electrostatic confinement in graphene is excluded 
by the Klein tunneling effect \cite{Klein}. However in bilayer graphene \cite{pereira,blqd} or buckled silicene \cite{ni} the energy gap
and thus the electrostatic confinement can be formed by perpendicular electric field.

In this paper we look for the counterpart of the Nagaoka ferromagnetism in the valley degree
of freedom in a 2D system. 
We focus on the ground-state valley polarization for a three-electron system in a square cluster of quantum dots
defined in silicene \cite{xenes}, i.e. a counterpart of the case studied experimentally in GaAs system \cite{naturexp}. 
With respect to the III-V quantum dots the silicene besides the valley degree of freedom  hosts
a strong intrinsic spin-orbit coupling \cite{soc,soc2} which as we show below plays a role in the Nagaoka ordering. 
Without the spin-orbit coupling term the tight-binding Hamiltonian is identical with the one
for the monolayer graphene  with staggered potential \cite{stag,stag2,stag3} up to the numerical value of the inter-atomic hopping energy. 
For that reason below we solve both the problems with and without the spin-orbit coupling.
We demonstrate that in the absence of the spin-orbit coupling the spin degree of freedom of the three-electron system needs to be frozen for the valley ordering to be observed. 
The intrinsic spin-orbit coupling splits
the fourfold degeneracy of the confined single-electron ground state with respect to the 
spin and valley forming spin-valley doublets in a manner similar to the one closely studied
for carbon nanotubes \cite{rmpcnt}. We demonstrate that in presence of the intrinsic spin-orbit coupling Nagaoka ordering in both valley and spin appear simultaneously. 
We discuss detection of spin-valley ordering by electron charge reaction to the sweep 
of confinement potentials in the cluster. 
Nagaoka ordering of the valley can be added to the toolbox of valleytronics \cite{valleytronics,ary}.

\section{Theory}

Below we study the system with the tight-binding approach \cite{soc,soc2,chow} and with the continuum approximation \cite{soc,EzawaNJP} to the tight-binding Hamiltonian.
The two approaches differ in the description of the valley degree of freedom, which is intrinsically
included in the Hamiltonian only in the continuum version, that neglects the intervalley scattering
by short-range deffects, the edges \cite{edge} of the flake and the short range component of the electron-electron interaction potential \cite{ivs1,ivs2,ivs3}. 
The diagonalization of the tight-binding Hamiltonian for localized states covers contribution of the entire Brillouin zone to the confined states
and the valley can only be resolved a posteriori. The atomistic method intinsically accounts for the
intervalley scattering. Application of the two methods allows for resolution of the intervalley scattering effects. The continuum method when applicable provides a radical reduction
of the numerical complexity with respect to the tight-binding (TB) approach. The latter
takes into orbitals localized on each atom while the nodes in the finite element  method (FEM) 
can be separated by much larger distances than depend only on the long range wave function
variation, so that the continuum approach can be applied to arbitrarily large systems.
However, due to the neglect of the intervalley scattering, the reliability of the 
FEM needs to be verified against the atomistic approach.

\subsection{Atomistic tight-binding Hamiltonian}

For the TB model we define a flake of buckled silicene \cite{hamsi,chow}
with ions of the A sublattice at positions ${\bf r}_{\bf k}^A=k_1 {\bf a}_1+k_2 {\bf a}_2$ with the crystal lattice vectors
${\bf a}_1=a \left(\frac{1}{2},\frac{\sqrt{3}}{2},0\right)$, 
and ${\bf a}_2=a \left(1,0,0\right)$, with silicene lattice constant $a=3.89$ \AA.
The B sublattice is shifted by a base vector ${\bf r}_{\bf k}^B={\bf r}_{\bf k}^A+(0,d,\delta)$ where $d=2.25$ \AA\; is the in-plane nearest-neighbor distance and the vertical distance is denoted by $\delta=0.46$ \AA.
Calculations are performed for a hexagonal flake with armchair edges and side length of about 30 nm with approximately 72 000 $p_z$ spin-orbitals.

We determine the eigenstates of atomistic TB Hamiltonian \cite{soc,soc2,chow}, 
\begin{eqnarray}
H_{TB}&=&-t\sum_{\langle k,l\rangle } p_{kl} c_{k}^\dagger c_{l}  +\sum_{k} V_k c^\dagger_{k}c_{k} \nonumber
\\ &+& it_{SO} \sigma_z\sum_{\langle \langle k,l\rangle \rangle  } p_{kl} \nu_{kl} c^\dagger_{k} c_{l}  
 +\frac{g\mu_B }{2}{\vec {\sigma}} \cdot {\bf{ B}}, \label{hb0}
\end{eqnarray}
where the first sum describes the nearest neighbor hopping with the energy $t=1.6$ eV \cite{soc,soc2}. In Eq. (\ref{hb0}) 
$p_{kl}=\exp\left({i\frac{e}{\hbar}\int_{\vec{r_k}}^{\vec{r_l}}\vec {\bf A}\cdot \vec {dl}}\right)$ stands for the Peierls phase.
 The integral in the exponent of $p_{kl}$ accounts for the Aharonov-Bohm phase shifts that the wave functions
 acquire from the vector potential ${\bf A}$ via hopping. We consider the magnetic field with both perpendicular $B_z$ and in-plane $B_x$ components ${\bf {B}}=(B_x,0,B_z)$ with the vector potential  ${\bf {A}}=(-B_z y/2, B_z x/2 -B_x z,0)$. 
Due to the 2D nature of the material
the in-plane component does not produce noticeable orbital effects. The in-plane field is introduced in order to manipulate spins of the confined states
 via the spin Zeeman effect included in the last term in Eq. (\ref{hb0}), with $\mu_B$ as the Bohr magneton and $g=2$ as the Land\'e factor.
 The third sum in Eq. (\ref{hb0}) introduces the intrinsic 
spin-orbit interaction \cite{km} with  the coupling constant $t_{SO}=3.9$ meV \cite{soc,soc2}
and  $\nu_{kl}=+1$  ($-1$)  for the path of the next-nearest neighbor hopping from ion $l$ to $k$ via the common neighbor that turns  counterclockwise (clockwise).   The second sum in Eq. (\ref{hb0}) introduces the external potential with $V_k$ standing for the potential on ${\bf r}_k$ ion.

In silicene the bias between sublattices opens the energy gap in the band structure \cite{ni} that allows
for formation of the confinement potential.  
For the confinement potential we assume that the bias is independent of the electron position within the plane,
\begin{equation} 
V_k=\left\{\begin{array}{cc} -V_g \sum_{i=1}^4 \left(\exp(-r_{ik}^2/R^2)-1\right) \text{on A}\\ -V_g \sum_{i=1}^4 \left(\exp(-r_{ik}^2/R^2)+1\right) \text{on B}
 \end{array}\right. \label{pott}
\end{equation}
where the sum over $i$ runs over 4 quantum dots with centers 
${\bf g}_i$ and ${\bf r}_{ik}=|{\bf r}_k-{\bf g}_i|$. We take $R=4.2$ nm for the dot radius,
and $V_g=0.3$ eV for the depth of potential cavities.
In Eq. (\ref{pott})  the electrostatic potential is lowered on both sublattices near the center of each quantum dot. 
This type of potential variation -- with nearly equal bias and a minimum on both sublattices --  can be achieved with a pair of flat gate electrodes with one that contains a circular
intrusion near the quantum dot center \cite{Zebrowski}. Note, that type I quantum dots
 can also be produced with flat gates
provided that they contain apertures near the confinement area \cite{scirep}.

Three-electron charge densities for the centers of the quantum dots ${\bf g}_i$
forming a square of side length $X$ are given in Fig. 1 with the A (B) sublattice placed on the left (right)
column of plots, and the side of the square increasing from top to bottom.

\begin{figure}[htbp]
\centering
\begin{tabular}{ll}
\includegraphics[width=0.22\textwidth]{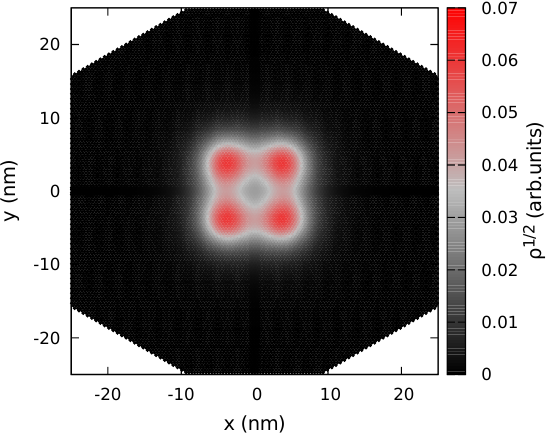} \put(-38,65){\color{white}(a)} & \includegraphics[width=0.22\textwidth]{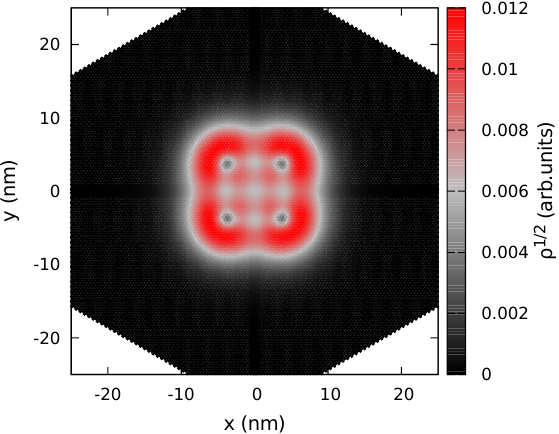}
\put(-38,65){\color{white}(b)}  \\
\includegraphics[width=0.22\textwidth]{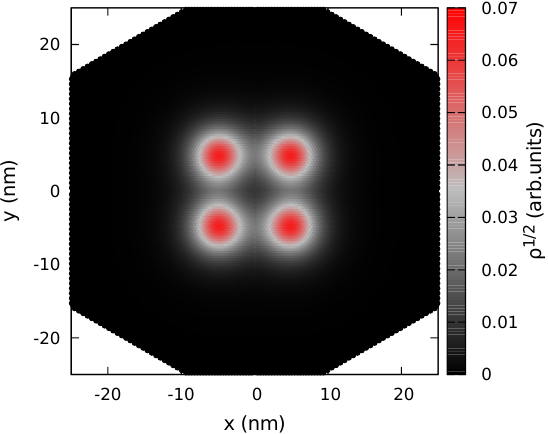} \put(-38,65){\color{white}(c)}  & \includegraphics[width=0.22\textwidth]{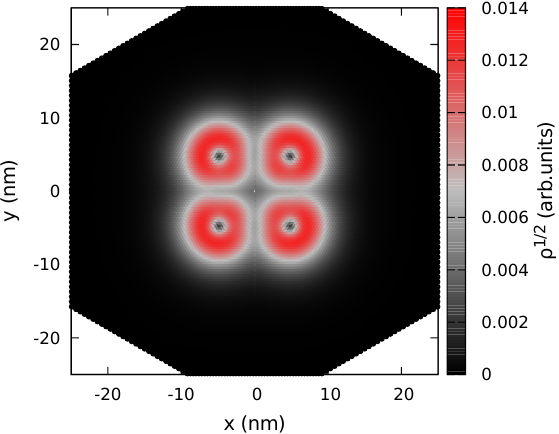}\put(-38,65){\color{white}(d)}   \\
\includegraphics[width=0.22\textwidth]{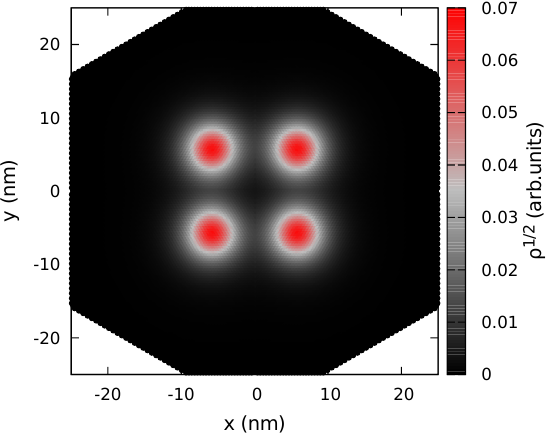} \put(-38,65){\color{white}(e)}  & \includegraphics[width=0.22\textwidth]{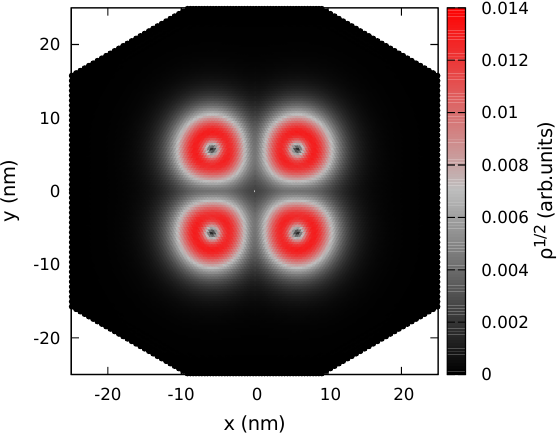}\put(-38,65){\color{white}(f)}  

\end{tabular}
\caption{Square-root of the charge density of the three-electron ground states as calculated with the TB 
 for neglected spin-orbit interaction.
The left (right) panels indicate the electron densities in the A (B) sublattice.
The rows of figures from top to bottom correspond to centers of the quantum dots placed on 
the corners of a square with side 8.4 nm (a-b), 10 nm (c-d), and 11.7 nm (e-f). 
}
\label{wf}
\end{figure}

\subsection{Continuum Hamiltonian}

The continuum approximation
explicitly resolves the valley degree of freedom. We work
with a four-component wave function spanned on sublattice and spin subspaces
$\psi=\left( \begin{matrix} \psi^{A\uparrow} & \psi^{B\uparrow} & \psi^{A\downarrow} & \psi^{B\downarrow}  \end{matrix}\right)^T$,
and the energy operator \cite{soc,EzawaNJP} 
\begin{eqnarray} 
H &=&\left[  \hbar v_F\left( k_x \tau_x -\eta k_y \tau_y \right)+ \left(\begin{array}{cc} V_A({\bf r}) & 0 \\ 0 & V_B({\bf r}) \end{array}\right)\right]\otimes {\bf I}_{spin} \nonumber  \\
&+& \eta t_{SO} \tau_z \otimes \sigma_z +{\bf I}_{sublattice} \otimes  \frac{g \mu_B} {2} \vec{\sigma}\cdot {\bf{B}} \nonumber \\
&-& W_D \nabla^2 \tau_z \otimes {\bf I}_{spin},
\label{con}
\end{eqnarray}
where $\sigma$ and $\tau$ are the Pauli matrices in the spin and sublattice subspaces, respectively.  ${\bf I}_{sublattice}$ and ${\bf I}_{spin}$  are the identity matrices,
 the Fermi velocity is $v_F=3dt/2\hbar$. The wave vector operators are defined as ${\bf k}=-i\nabla +\frac{e}{\hbar}{\bf A}$, and $\eta=\pm 1$ is the valley index. 

Hamiltonian (\ref{con}) is diagonalized by the FEM. The computational box is divided into typically about 2200 triangular elements with 18000 nodes supporting
Lagrange interpolating polynomials of the second degree \cite{Solin} as shape functions covering the spin and sublattice spaces.
In Eq. (\ref{con}) the last 
 expression is an artificial Wilson term \cite{wi} that is applied to remove the spurious states \cite{wi,spur2,spur3,spur4} due to the fermion doubling problem from the low-energy spectrum.
We take  the Wilson parameter $W_D=36$ meV nm$^2$ which increases the energy of the fast oscillating states with a negligible influence on the actual solutions of the Dirac equation that are smooth near the charge neutrality point.

\subsection{Calculations for three electrons}

The electron-electron interaction for gapless graphene flakes leads to generation of   electron and hole pairs \cite{egger}.
In the calculations that follow for the three-electron system the typical total interaction energy is about 60 meV, i.e. $\simeq 20$ meV per electron pair, i.e. much lower than the 
potential bias between the sublattices.
Since the interaction energy is lower than the field-induced energy gap and the considered quantum dot does not
support confinement of holes we neglect the effects of pair generation by Coulomb interaction
\cite{egger} and assume that the number of conduction band electrons
is fixed \cite{c2}. 
In both the atomistic method and in FEM we diagonalize the Hamiltonian
in the basis of three-electron wave functions constructed by the lowest-energy 48 confined eigenstates   of the conduction band that produces the basis of 17 296 Slater determinants.

The Hamiltonian for the system of interacting electrons is
\begin{equation}
  {H}_{i}=\sum_{i}{d}^{\dagger}_{i}{d}_{i}E_i +\frac{1}{2}\sum_{ijkl}{d}^{\dagger}_{i}{d}^{\dagger}_{j}{d}_{k}{d}_{l}V_{ijkl},
  \label{Htwo}
\end{equation}
where ${d}^{\dagger}_{i}$ is the electron creation operator for the energy level $E_i$.
The two-electron Coulomb matrix elements that 
\begin{equation}
V_{ijkl}=\kappa\langle{\psi_{i}(\mathbf{1})\psi_{j}(\mathbf{2})|\frac{1}{|\mathbf{r_{12}}|}}|{\psi_{k}(\mathbf{1})\psi_{l}(\mathbf{2})}\rangle, 
\label{coulF}
\end{equation}
with $\kappa=e^2/(4\pi\epsilon\epsilon_0)$. We take $\epsilon_0=4.5$ for the dielectric constant, that corresponds to SiO$_2$ or thin layers of Al$_2$O$_3$ \cite{al2o3groner,birey}
applied as a matrix embedding the silicene monolayer.

\subsubsection{Coulomb integrals in the continuum approach}

For the continuum approach we calculate the Coulomb matrix elements using the formula
\begin{eqnarray}
V_{ijkl}&=&\kappa\delta_{\eta_i,\eta_k}\delta_{\eta_j,\eta_l}  \int \int d{\bf r}_1 d{\bf r}_2 
\left[\psi_{i}^{A\uparrow*}({\bf r}_1)\psi_{k}^{A\uparrow}({\bf r}_1) \right. \nonumber \\
&+&\psi_{i}^{B\uparrow*}({\bf r}_1)\psi_{k}^{B\uparrow}({\bf r}_1) 
+\psi_{i}^{A\downarrow*}({\bf r}_1)\psi_{k}^{A\downarrow}({\bf r}_1) \nonumber \\
&+&\left. \psi_{i}^{B\downarrow*}({\bf r}_1)\psi_{k}^{B\downarrow}({\bf r}_1)\right]
\frac{1}{|\mathbf{r_{12}}|} \left[\psi_{j}^{A\uparrow*}({\bf r}_2)\psi_{l}^{A\uparrow}({\bf r}_2)\right. \nonumber \\
&+&\psi_{j}^{B\uparrow*}({\bf r}_2)\psi_{l}^{B\uparrow}({\bf r}_2) 
+\psi_{j}^{A\downarrow*}({\bf r}_2)\psi_{l}^{A\downarrow}({\bf r}_2) \nonumber \\
&+&\left. \psi_{j}^{B\downarrow*}({\bf r}_2)\psi_{l}^{B\downarrow}({\bf r}_2)\right]. \label{xx}
\end{eqnarray}
The deltas with the valley indices that stand before the integral imply the neglect of the inter-valley 
scattering effects \cite{ivs1,ivs2,ivs3} that are accounted for 
only in the TB approach.

\subsubsection{Coulomb integrals in the atomistic approach}
In the TB method the single-electron wave functions $\psi$ are expanded
in the basis of $3p_z$ spin-orbitals of Si ions,
\begin{equation}
\psi_i({\bf r}_i)=\sum_{k,\sigma_k} C^i_{k,\sigma_k} p_z^k({\bf r}_1).
\end{equation}
The Coulomb matrix elements are summed over the ions,
\begin{eqnarray}
V_{ijkl}&=&\kappa\sum_{\substack{a,\sigma_{a};b,\sigma_{b};\\ c,\sigma_{c};d,\sigma_{d} }}C_{a,\sigma_{a}}^{i*}C_{b,\sigma_{b}}^{j*}C_{c,\sigma_{c}}^{k}C_{d,\sigma_{d}}^{l}\delta_{\sigma_{a};\sigma_{d}}\delta_{\sigma_{b};\sigma_{c}}\times \nonumber\\&&\langle p_{z}^{a}({\bf r}_{1})p_{z}^{b}({\bf r_{2}})|\frac{1}{|{r_{12}}|}|p_{z}^{c}({\bf r_{1}})p_{z}^{d}({\bf r_{2}})\rangle.
\end{eqnarray}
In the two-center approximation \cite{c2} 
 $ \langle p_{z}^{a}({\bf r}_{1})p_{z}^{b}({\bf r_{2}})|\frac{1}{|{r_{12}}|}|p_{z}^{c}({\bf r_{1}})p_{z}^{d}({\bf r_{2}})\rangle=\frac{1}{r_{ab}} \delta_{ac}\delta_{bd}$ for $a\neq b$. 
The  on-site integral ($a=b$) for $3p_z$  Si atomic orbitals with  $p_z({\bf r})=N z \left(1-\frac{Z r}{6}\right)\exp(-Zr/3)$, where $N$ stands for the normalization  and $Z$ is the effective screened nucleus charge equals $\frac{3577}{46080}Z$. The Slater screeing rules for $3p$ Si orbitals gives $Z=4.15$.

\begin{figure}[htbp]
\centering
\begin{tabular}{ll}
\includegraphics[width=0.17\textwidth]{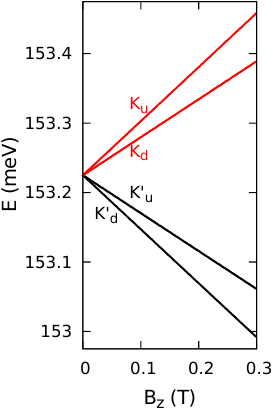} \put(-17,50){(a)} & \includegraphics[width=0.17\textwidth]{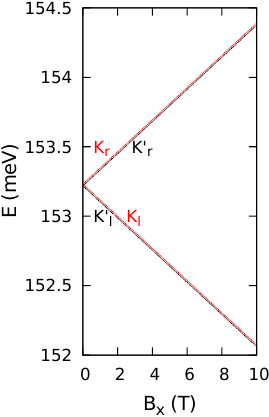}
\put(-17,50){(b)}  \\
\includegraphics[width=0.17\textwidth]{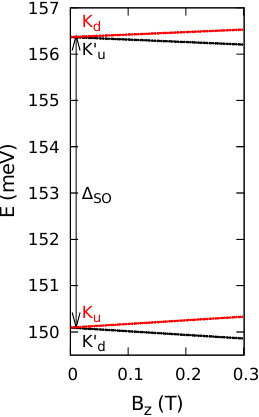} \put(-17,40){(c)} & \includegraphics[width=0.17\textwidth]{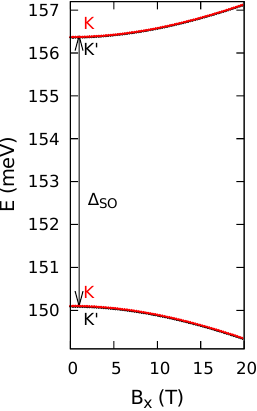}\put(-17,40){(d)}  \put(-52,47) {\includegraphics[width=0.08\textwidth]{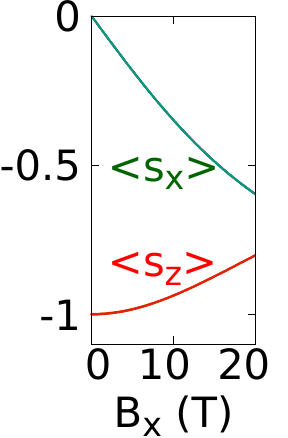} }
\end{tabular}
\caption{Low-energy single-electron spectrum of states confined in a {\it single} quantum dot
without (a-b) and with (c-d) the spin-orbit  interaction. The results in (a,c) 
are obtained for $B_x=0$ and in (b,d) for $B_z=10$mT. The subscripts $u/d$ in (a,c) correspond to spin-up and spin-down eigenstates of $\sigma_z$ operator. In (b) $l$, and $r$ subscripts stand for the eigenstates of $\sigma_x$ operator with negative and positive eigenvalue, respectively.
The inset in (d) shows the average values of $\sigma_x$ and $\sigma_z$ operators for the ground-state.}
\label{piery}
\end{figure}

\begin{figure}[htbp]
\centering
\begin{tabular}{ll}
 \includegraphics[width=0.22\textwidth]{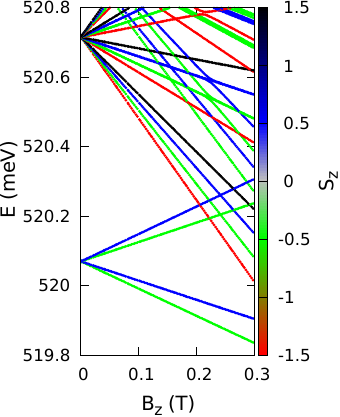}  \put(-55,24){(a)} &
 \includegraphics[width=0.22\textwidth]{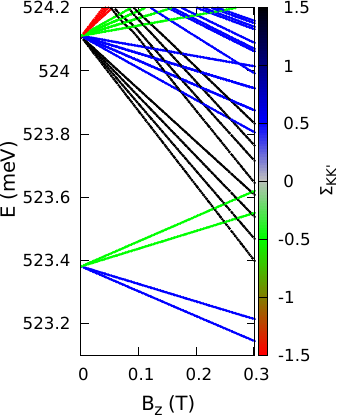} 
 \put(-55,24){(b)} \\
 \includegraphics[width=0.22\textwidth]{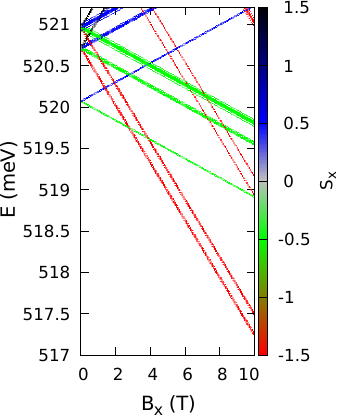} \put(-55,24){(c)} 
 & \includegraphics[width=0.22\textwidth]{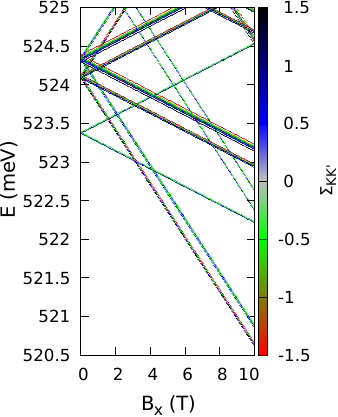}\put(-55,24){(d)} 

\end{tabular}
\caption{Three-electron energy levels for the quadruple quantum dot as calculated within 
TB (a,c) and FEM (b,d). The results in (a,c) 
are obtained for $B_x=0$ and in (b,d) for $B_z=10$mT. 
 Spin-orbit interaction is neglected ($t_{SO}=0$). 
The color scale in (a,c) and (b,d) indicate the z-component of the spin and the valley isospin component, respectively. 
The centers of the four quantum dots are placed on the corners of a square with side length of $X=11.7$ nm.
}
\label{3els}
\end{figure}

\section{Results and discussion}
\subsection{Single-dot single-electron results}
In the absence of spin-orbit interaction and without external magnetic field the single-electron confined ground state in a single quantum dot
is four-fold degenerate with respect to both spin and valley [Fig. \ref{piery}(a)]. 
For $B_z=0$ the valley degeneracy
is preserved for the in-plane field $B_x$ and the energy levels are split only 
with respect to the spin. In Fig. \ref{piery}(b) and in other plots
of this work presented as functions of $B_x$ we apply a residual perpendicular magnetic field
for $B_z=10$mT which lifts the degeneracies of the energy levels to a thickness of a line.

The intrinsic spin-orbit coupling introduces an effective magnetic field perpendicular 
to the plane of confinement with orientation
that is opposite in the sense of eigenvalue sign to the valley isospin $\eta$ (cf. the term with $t_{SO}$ coupling constant
in Eq. (\ref{con})).  
This effective field splits [Fig. \ref{piery}(c)] the ground state into a pair
of doublets with  splitting energy that corresponds to the spin Zeeman effect at the magnetic field as large as $\sim 34$ T. 
 Application of an in-plane magnetic field [Fig. \ref{piery}(d)] slowly tilts the spins to the  $x$ direction [see the inset to Fig. \ref{piery}(d)].

\begin{figure}[htbp]
\centering
\begin{tabular}{ll}
\includegraphics[width=0.22\textwidth]{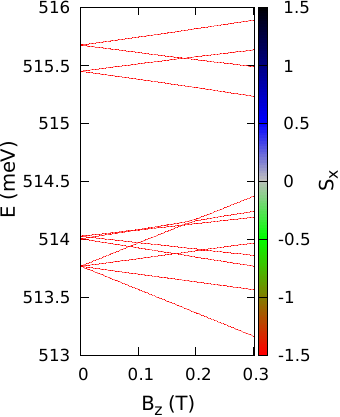}\put(-41,75){(a)} &
\includegraphics[width=0.22\textwidth]{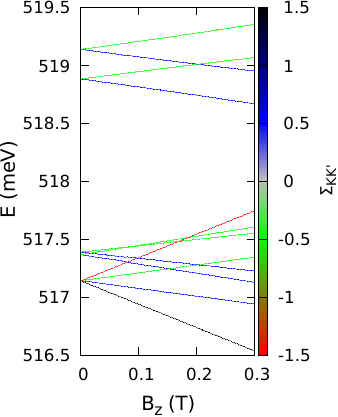} \put(-41,75){(b)} 
\end{tabular}
\caption{Same as Fig. \ref{3els}(a,b) only with in-plane magnetic field $B_x=20$ T.
(a) shows the TB results  (b) the FEM ones.
The centers of the four quantum dots are placed on the corners of a square with side length of $X=11.7$ nm and $t_{SO}=0$. }
\label{w20bxnoso}
\end{figure}

\subsection{Three electrons in quadruple quantum dot for $t_{SO}=0$}

For a  single circular quantum dot the single-electron ground state corresponds to angular momentum quantum number 0 
for the wave function component on sublattice A and $\pm 1$ on sublattice B \cite{Zebrowski}.
In consequence, the charge density in a single quantum dot corresponds to a  maximum and a zero of the charge density in the centers of the quantum dots on A and B sublattice, respectively.
The single-electron properties are consistent with the charge density distribution for three interacting electrons 
with well separated wave functions -- see  Fig. \ref{wf}(e,f) for the  centers of potential
minima distributed on a square with side length of $X=11.7$ nm. 
The wave function component on the B sublattice is less strongly localized and thus it mediates the inter-dot tunneling in a stronger extent.

\subsubsection{Nagaoka valley ordering}

The three-electron spectrum is given in Fig. \ref{3els}(a) (TB) and in Fig. \ref{3els}(b) (FEM). 
The results of the two approaches agree very well up to a relative shift 
of the entire spectra on the energy scale of a few meV. In the results of the atomistic approach
we plot the energy levels with colors indicating  the total spin $z$ or $x$ component. On the energy levels calculated with  FEM we  mark by the color of the lines the total valley isospin component for the three electrons. 

The three-electron ground-state at $B_z=0$ is four-fold degenerate. The degenerate
energy levels correspond to the eigenvalue of $S_z=\frac{1}{2}\sum_{i=1}^3 \sigma^i_z$ component of the total spin $\pm \frac{1}{2}$ and the $z$-component of the total valley isospin of $\Sigma_{KK'}=\frac{1}{2}\sum_{i=1}^3 \eta^i_z$ equal to $\pm \frac{1}{2}$. 
The ground-state is not polarized neither in spin nor in valley. The first excited state is sixteen-fold degenerate. The energy levels that are degenerate at $B_z=0$ correspond to the total valley index $\Sigma_{KK'}=-\frac{3}{2},-\frac{1}{2},\frac{1}{2},\frac{3}{2}$
and the z-components of the spin $S_z=-\frac{3}{2},-\frac{1}{2},\frac{1}{2},\frac{3}{2}$.

Nagaoka ferromagnetism was observed \cite{naturexp} in a quadruple quantum dot defined in GaAs, with three electrons per 8 available spin-orbitals. 
 In silicene system we have 16 available spin-valley-orbitals due to the additional valley degree of freedom. One can try to eliminate the spin degree of freedom and reduce the number of equivalent states to 8  by applying a strong magnetic field to freeze the spin degree of freedom. The application of the perpendicular magnetic field lifts the valley degeneracy [Fig. \ref{piery}(a)] that would eventually lead to 
 the valley polarization by the external field. Here, we want to preserve the valley degeneracy 
in order to study the electron-electron interaction triggering the valley polarization. For that reason we choose to apply the in-plane field that interacts only with the spin and not the valley of confined states [Fig. \ref{piery}(b)]. The results are given in Fig. \ref{3els}(c,d) with a residual $B_z=10$ mT applied to slightly split the energy levels for visualization. The states  that are spin polarized in the $-x$ direction for all valley configurations are promoted to the lower energy by the $B_x$ field.

The structure of the low-energy spectrum spin-polarized by strong $B_x$ field is revealed once
a weak $B_z$ field is additionally applied. In Figure \ref{w20bxnoso} we set $B_x=20$ T and calculate
the energy levels as functions of $B_z$. In the applied range of $B_z\leq 0.3$ T the spins remain nearly perfectly polarized in the $-x$ direction. The ground-state at $B_z=0$ is fourfold degenerate with the valley isospin component that takes  values $-\frac{3}{2},-\frac{1}{2},\frac{1}{2}$ and $\frac{3}{2}$.  This structure of the
ground-state energy level is a valley-ordered counterpart of spin-polarized three-electron state with the total spin quantum number $S=\frac{3}{2}$.
In the excited part of the spectrum in Fig. \ref{w20bxnoso} we find a number of valley non-polarized states forming doublets at $B_z=0$ with the valley isospin equal to $\pm \frac{1}{2}$. 
By analogy to the spin degree of freedom we attribute
the total valley isospin quantum number $V=\frac{3}{2}$ to the four-fold degenerate state with the 
isospin component $\Sigma_{KK'}$ changing from $-V$ to $V$ with steps of 1. The doublets thus correspond to $V=\frac{1}{2}$ with the components  $\Sigma_{KK'}=\pm \frac{1}{2}$. 

\begin{figure}[htbp]
\centering
\includegraphics[width=0.35\textwidth]{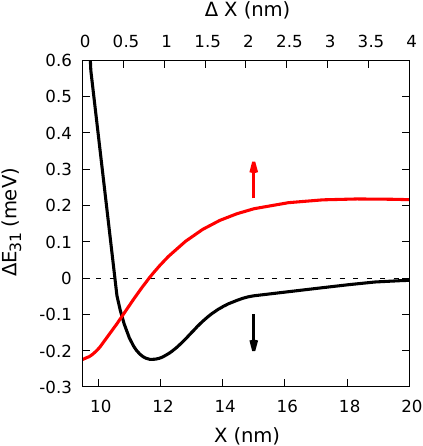} \put(-119,119) {\includegraphics[width=0.1\textwidth]{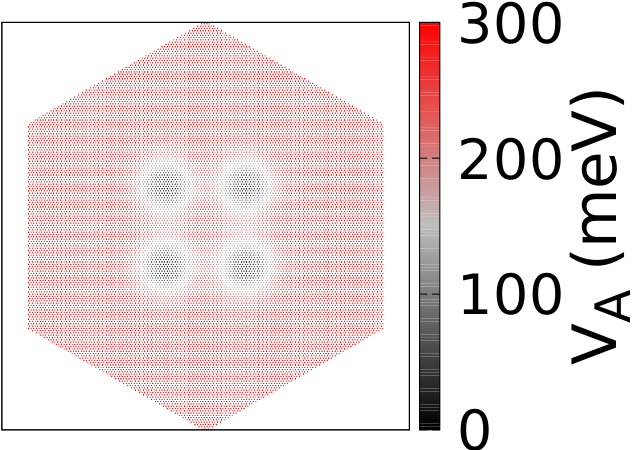} 
}
 \put(-52,121) {\includegraphics[width=0.065\textwidth]{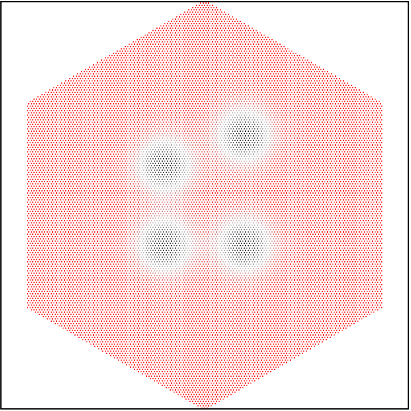} }
\caption{ The black lines show the energy difference (FEM) between the lowest valley polarized and non-polarized states for $B_x=20$ T and $t_{SO}=0$ as a function of the side length of a square on which the centers of the four quantum dots are placed. The red line shows the results obtained for $X=11.7$ nm, that corresponds to the most stable valley-polarized ground state as a function of a shift $\Delta x$
of the $y$ position of the quantum dot localized in the first quadrant of the Cartesian coordinate.
The insets show the potential on the A sublattice for $X=11.7$ nm and $\Delta X=0$ (left)
and $\Delta X=4$ nm (right). Same scale for $V_A$ is applied for both sublattices. 
The frame in the insets has the length of 60 nm. 
}
\label{faz}
\end{figure}

The Nagaoka polarization of the ground-state appears due to the inter-dot
electron tunneling and thus it is determined by the system geometry.  
In order to study the valley polarization we kept $B_x=20$T and varied the 
positions of the centers of the dots. 
We define $\Delta E_{31}$ as the energy difference between the lowest valley polarized state with $V=\frac{3}{2}$ and the lowest non-polarized state with $V=\frac{1}{2}$ state. $\Delta E_{31}<0$ corresponds to the Nagaoka ordered valley in the ground-state. The black line in Fig. \ref{faz} shows the result as a function
of $X$ -- the side length of the square on which the centers of the dots are placed [Fig. \ref{wf}]. The valley-polarized ground-state
is found for $X\gtrsim 10.5$nm. The polarized ground-state is most stable for $X=11.7$nm that was selected for plots in Fig. \ref{wf}(e,f), Fig. \ref{3els} and Fig. \ref{w20bxnoso}.
For low values of $X$ the four quantum dots become
more strongly tunnel coupled and they eventually are transformed to a single quantum dot 
for which the ground-state is not polarized. In the limit of large $X$
the inter-dot tunneling is negligible. Once the electrons are separated and their hopping between
the dots is removed the valley isospin has no influence on the energy for $B_z=0$, hence the degeneracy of $V=\frac{3}{2}$ and $V=\frac{1}{2}$ states at large $X$. 

Ref. \cite{naturexp} found that the ground-state spin ordering vanishes when the array of dots is deformed
to approach the limit of a quantum dot chain. For a chain of dots  \cite{Bd1} the Lieb-Mattis theorem \cite{LM1,LM2} excludes  spin polarization of the ground state. Here we looked for a similar effect
in the valley degree of freedom. We fixed the value of $X$ to 11.7 nm and 
 moved the quantum dot localized in the first quadrant of the coordinate system ($x>0,y>0$,
see the insets to Fig. \ref{faz}) shifting its center by $\Delta X$ in the $y$ direction. The results
are displayed in Fig. \ref{faz} by the red line. The shift by $\Delta X \simeq 0.75$ nm
makes the valley polarized and non-polarized states degenerate. For $\Delta X\gtrsim 2.1$ nm 
the value of $\Delta E_{31}$ is more or less inverted from the $\Delta X=0$ case.

\subsubsection{Inter-valley scattering effects}
The inter-valley scattering can appear as (i) a single-electron effect triggered by the armchair edge \cite{edge} of the flake and (ii) as an interaction effect due to the short-range component of the Coulomb potential \cite{ivs1,ivs2,ivs3}. 

\begin{figure}[htbp]
\centering
\includegraphics[width=0.15\textwidth]{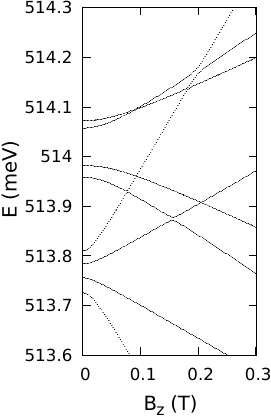} \put(-32,105){(a)}
\includegraphics[width=0.15\textwidth]{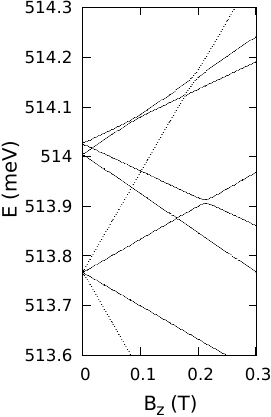} \put(-32,105){(b)}
\includegraphics[width=0.15\textwidth]{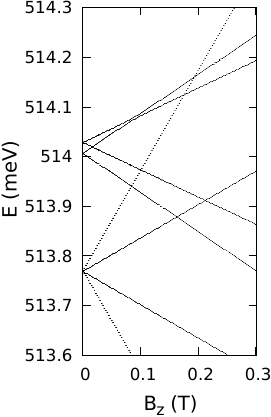} \put(-32,105){(c)}
\caption{Enlarged low-energy part of the spectra for parameters of Fig. \ref{w20bxnoso} 
for the side length of the hexagonal silicene flake of length 20.5 nm (a), 23 nm (b), and 25 nm (c)
as calculated with the atomistic method.}
\label{zoom}
\end{figure}

The effect (i) can be observed for a smaller flake, when the tails of quantum-dot-confined wave functions tunnel to the armchair edge that induces inter-valley mixing. 
 Figure \ref{zoom} shows a zoom of the low-part of the spectra for parameters applied in Fig. 4 with the side length of the flake of 20.5 nm [Fig. \ref{zoom}(a)], 23 nm [Fig. \ref{zoom}(b)] and 25 nm [Fig. \ref{zoom}(c)]. 
The valley mixing due to the  edge effect lifts the degeneracy at $B_z=0$ and opens avoided crossings between
energy levels that in the continuum approach correspond to different valley isospin quantum numbers [compare with Fig. 4(b)]. For strong inter-valley mixing the dependence of energy
levels of $B_z$ deviates from linear, in particular near $B_z=0$.

\begin{figure}[htbp]
\centering
\includegraphics[width=0.22\textwidth]{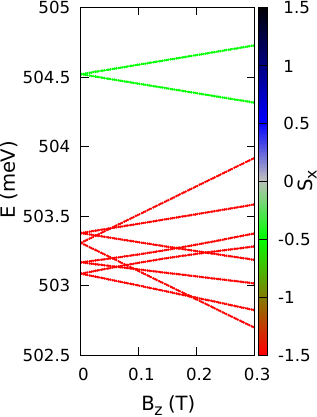} \put(-37,135){(a)}\includegraphics[width=0.22\textwidth]{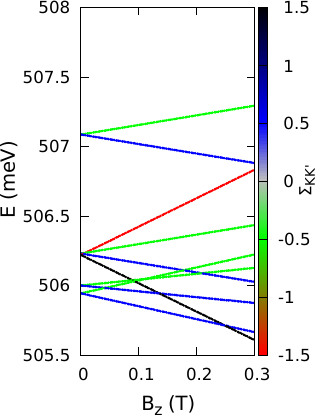} \put(-37,135){(b)}
\caption{Same as Fig. \ref{w20bxnoso} only with the inter-dot distance decreased to $X=10$ nm. }
\label{closer}
\end{figure}

\begin{figure}[htbp]
\centering
\includegraphics[width=0.215\textwidth]{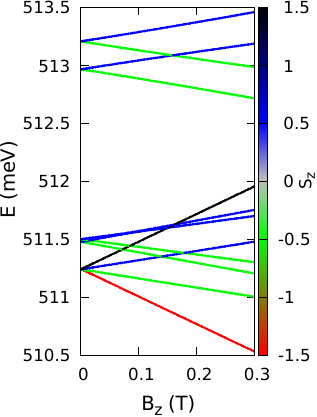} \put(-37,95){(a)}
\includegraphics[width=0.23\textwidth]{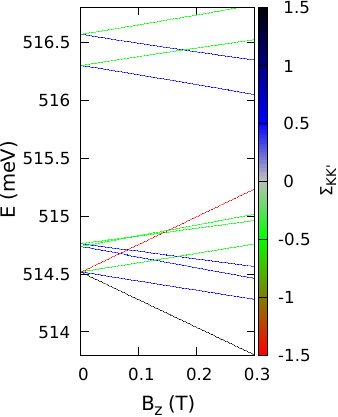} \put(-47,95){(b)}
\caption{Same as Fig. \ref{w20bxnoso} only with $t_{SO}=3.9$ meV,  $B_x=0$.
Color scale in (a) now shows the $z$ component of the spin.}
\label{zso}
\end{figure}

The effect (ii) is not triggered when the electrons occupy separate quantum dots. In this case
only the long range tail of the Coulomb potential is resolved by carriers.

In Fig. \ref{closer} we plotted the results for quantum dot centers placed
at the corners of the square of side length $X=10$ nm. For this parameters the valley ordering is no longer observed in the ground-state [see Fig. \ref{faz}]. 
In Fig. \ref{closer} we can see that the agreement between the two methods is no longer
as perfect as above
In particular, the valley-ordered quadruplet is  found here
only by the continuum approach but in the TB the quadruplet is split into two doublets.
The two-electron levels splitting by the inter-valley scattering due to the electron-electron interaction  was discussed in detail in Ref. \cite{Zebrowski} for an electron pair in a single quantum dot (see Fig 4(a) and Fig. 4(b) in Ref. \cite{Zebrowski}). 

We conclude that in the parameter range where Nagaoka ordering  is found the inter-valley 
scattering by the Coulomb interaction is negligible or missing for system geometry
for which Nagaoka valley ordering appears.

\subsection{Nagaoka polarization in presence of the spin-orbit interaction}

Figure \ref{zso} shows the spectra for $X=11.7$ nm and $t_{SO}=3.9$ meV in the absence
of the in-plane field $B_x=0$. 
The pattern of energy levels is similar to the one found
in Fig. \ref{w20bxnoso} for $t_{SO}=0$ and $B_x=20$ T. 
For $t_{SO}=3.9$ meV and $B=0$ a single dot hosts
a two-fold degenerate ground state [Fig. \ref{piery}(c,d)] with opposite spin and valley isospin components
$\eta \sigma_z=-1$. 
This pattern of energy levels replaces the 
valley-degenerate ground-state with $\eta=\pm 1$ found for $t_{SO}=0$, $B_z=0$ and strong in-plane field that freezes the spin found for Fig. \ref{piery}(b).
In these two above cases the Coulomb integrals between the single-electron 
states in the low-energy part of the spectrum are similar, hence the agreement
of the three-electron spectra in Fig. \ref{zso} and Fig. \ref{w20bxnoso}.
In presence of the intrinsic spin-orbit interaction there is a perfect anticorrelation between the $S_z$ and $\Sigma_{KK'}$ quantum numbers  [Fig. \ref{zso}(a,b)]. All the information on the eigenstates is therefore redundantly included in 
the spin and valley sets of quantum numbers. 

\subsection{Detection of polarized states}

The confined spectra including the low-energy excited states can be studied
with the transport spectroscopy that was developped for GaAas quantum dots \cite{ts1}
and more recently applied to graphene quantum dots \cite{ts2,ts3} or carbon nanotubes
\cite{ts4,rmpcnt}. The spin-valley structure of the spectrum can be extracted
from the degeneracy and slopes of the energy levels \cite{rmpcnt,ts5} in external magnetic field that can be measured with a precision of several $\mu$eV \cite{ts5}.
Besides the energy level dependence on the external magnetic field
the detection of the ground-state spin polarization is performed in experiments
on multiple quantum dots using the confinement potential variation \cite{petta,naturexp,sitrisi}.
The confinement potential is first fixed for a time long enough for the 
electron system to relax to the ground state. Next the potential undergoes a change \cite{petta,naturexp,sitrisi}
such that after the sweep the ground-state corresponds to two electrons in one of the dots.
The change is applied diabatically, i.e. faster than the spin or valley relaxation time.
The ground state with double occupancy of a quantum dot can only occupied provided
that initially the electrons are not polarized, otherwise
the double occupancy is forbidden by Pauli exclusion (valley-spin blockade \cite{ts4}).
The charge of the dots is monitored by on-chip charge sensors \cite{petta,naturexp,sitrisi}. 

For the study of charge redistribution we take the system considered in the precedent
subsection and generalize the potential of Eq. (\ref{pott}) in order to cover
confinement variation
\begin{equation} 
V_k({\bf r}_k)=\left\{\begin{array}{cc} -V_g \sum_{i=1}^4 \left(\exp(-r_{ik}^2/R^2)-1\right)/\alpha_i \;\text{on A}\\ -V_g \sum_{i=1}^4 \left(\exp(-r_{ik}^2/R^2)+1\right)/\alpha_i \;\text{on B}
 \end{array}\right.,\label{pottg}
\end{equation}
where we take $\alpha_i=\alpha$ for the three dots localized at $x<0$ or $y<0$ leaving
$\alpha_i=1$ for the dot of the first quadrant. 

\begin{figure}[htbp]
\begin{tabular}{l}
\includegraphics[width=0.3\textwidth]{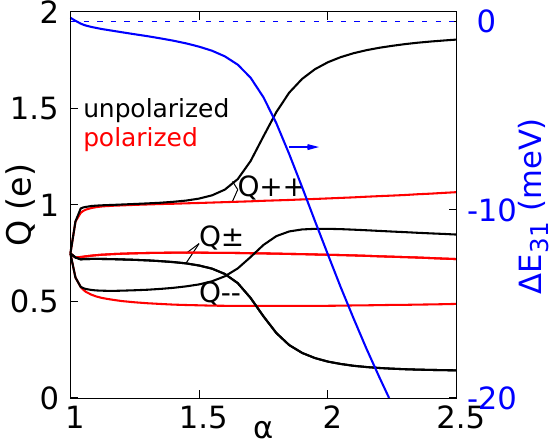} \put(0,110){(a)} \\
 {\includegraphics[width=0.1\textwidth]{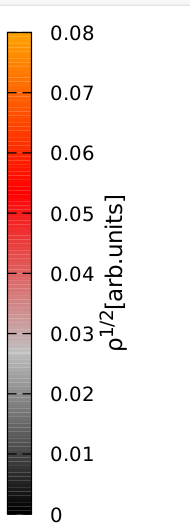}}  {\includegraphics[width=0.25\textwidth]{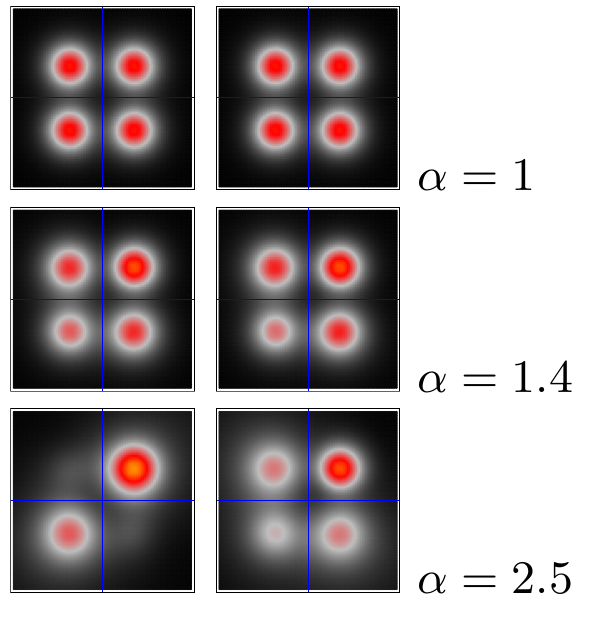}}  \put(-30,110){(b)}
\end{tabular}
\caption{
(a) Charge localized in the upper-right ($x>0,y>0$) quadrant ($Q$++), the lower-left ($x<0,y<0$) quadrant $Q--$ and in the other quadrants $Q\pm$ $(x<0,y>0)$ or $(x>0,y<0)$ as function of the divisor $\alpha$
which lowers the confinement potential in the left and lower quadrants $x<0$ or $y<0$ see Eq. (\ref{pott}) for the lowest-energy spin-valley polarized state (red lines)
and the lowest-energy unpolarized state (black lines). 
The blue line shows the energy (FEM) difference between the polarized and unpolarized levels ($\Delta E_{31}$ right axis).
(b) Square root of the charge density for the lowest polarized and unpolarized energy levels. The side of the each square plot is 34 nm long.
Parameters are the same as in Section III.C. 
}
\label{charge}
\end{figure}

The charge localized in each quadrant for the lowest-energy 
spin-valley polarized state and the lowest-energy unpolarized state for $B=0$
are displayed in Fig. \ref{charge}(a) as a function of $\alpha$.
The square of the charge density is plotted in Fig. \ref{charge}(b).
For $\alpha=1$ we have 3/4 electron charge per quantum dot in both polarized
and unpolarized state. A difference in the charge distribution can only appear
when the potential symmetry is lifted.
 As $\alpha$ is increased from 1 the dots
on the left and lower side of the cluster are made shallower.
The reaction of the charge in both the states is at first similar as $\alpha$
is increased from 1. The dot upper-right dot captures an entire electron charge
at the expense of the other dots.
Moreover, the unpolarized state becomes the ground-state for $\alpha\geq1.028$.
Polarization removal from the ground-state is consistent with
the results for the deformed cluster of dots (see Fig. \ref{faz}). 
As $\alpha$ is increased further in the unpolarized state a second
electron starts to occupy the dot of the first quadrant and simulatenously the charge
of the opposite dot is increased to minimize the inter-dot electron-electron interaction
energy. A double occupancy of the dot is forbidden for the polarized state
and its charge distribution does not change much when $\alpha$ is increased
above 1.1. In order to detect the initial  spin-valley polarization
one needs to diabatically sweep $\alpha$ from 1 to i.e. 2.5 and next
measure the charge localized in the dot of the first quadrant. 
The procedure should also be useful for determination of the spin and valley relaxation times.

 \section{Summary and conclusions}
We studied a system of three-electrons in a square cluster of quantum dots defined within material that provides valley degree of freedom to the confined single-electron states using the continuum approach
that allows for identification of the valley isospin in the atomistic TB spectra. 
We found that the Nagaoka-type polarization of the valley in a system without the intrinsic spin-orbit coupling is found in conditions when the spin degree of freedom is frozen by in-plane magnetic field. 
Non-polarized ground state is promoted when the spatial symmetry of the cluster is lifted by
a shift of one of the quantum dots that transforms the cluster toward a chain of quantum dots. 
In presence of the intrinsic spin-orbit coupling the spin-valley polarization is observed 
along with the perfect anticorrelation of the spin and valley isospin components 
already in the absence of external magnetic field. 
The pattern of the energy levels near the ground-state for systems with and without the spin-orbit coupling is very similar provided that a strong in-plane magnetic field is applied to the latter.
The spontaneous ground-state valley polarization  
in the system can be harnessed for studies of valley manipulation in multiple quantum dots.
The Nagaoka valley polarization can be detected by charge conversion
using the Pauli blockade of the double occupancy of a quantum dot.

\section*{Acknowledgments}
This work was supported by the National Science Centre
(NCN) according to decision DEC-2016/23/B/ST3/00821.
Calculations were performed on PLGrid infrastructure.

\end{document}